\documentclass[aps, prd, twocolumn, superscriptaddress, nofootinbib]{revtex4}
\usepackage{graphicx}
\usepackage{dcolumn}
\usepackage{amssymb}
\usepackage{amsmath,amssymb,amsfonts}

\begin{document}

\newcommand{\Eq}[1]{\mbox{Eq. (\ref{eqn:#1})}}
\newcommand{\Fig}[1]{\mbox{Fig. \ref{fig:#1}}}
\newcommand{\Sec}[1]{\mbox{Sec. \ref{sec:#1}}}

\newcommand{\PHI}{\phi}
\newcommand{\PhiN}{\Phi^{\mathrm{N}}}
\newcommand{\vect}[1]{\mathbf{#1}}
\newcommand{\Del}{\nabla}
\newcommand{\unit}[1]{\;\mathrm{#1}}
\newcommand{\x}{\vect{x}}
\newcommand{\y}{\vect{y}}
\newcommand{\p}{\vect{p}}
\newcommand{\ScS}{\scriptstyle}
\newcommand{\ScScS}{\scriptscriptstyle}
\newcommand{\xplus}[1]{\vect{x}\!\ScScS{+}\!\ScS\vect{#1}}
\newcommand{\xminus}[1]{\vect{x}\!\ScScS{-}\!\ScS\vect{#1}}
\newcommand{\diff}{\mathrm{d}}
\newcommand{\mk}{{\mathbf k}}
\newcommand{\ep}{\epsilon}

\newcommand{\be}{\begin{equation}}
\newcommand{\ee}{\end{equation}}
\newcommand{\bea}{\begin{eqnarray}}
\newcommand{\eea}{\end{eqnarray}}
\newcommand{\vu}{{\mathbf u}}
\newcommand{\ve}{{\mathbf e}}
\newcommand{\vn}{{\mathbf n}}
\newcommand{\vk}{{\mathbf k}}
\newcommand{\vz}{{\mathbf z}}
\newcommand{\vx}{{\mathbf x}}
\def\dup{\;\raise1.0pt\hbox{$'$}\hskip-6pt\partial\;}
\def\ddn{\;\overline{\raise1.0pt\hbox{$'$}\hskip-6pt\partial}\;}


\title{Correlation between opposite-helicity gravitons: Imprints on gravity-wave and microwave backgrounds}

\newcommand{\addressImperial}{Theoretical Physics, Blackett Laboratory, Imperial College, London, SW7 2BZ, United Kingdom}
\newcommand{\addressRoma}{Dipartimento di Fisica, Universit\`a ‚ÄúLa Sapienza‚Äù
and Sez. Roma1 INFN, P.le A. Moro 2, 00185 Roma, Italia}

\author{Giulia Gubitosi}
\affiliation{\addressImperial}
\author{Jo\~{a}o Magueijo}
\affiliation{\addressImperial}

\date{\today}

\begin{abstract}
We examine some of the roots of parity violation for gravitons and uncover a closely related new effect: 
correlations between right and left handed gravitons. Such correlators have spin 4 if they involve gravitons moving along the same direction,
and spin zero for gravitons moving with opposite directions. In the first case, the most immediate implication would be a degree 
of linear polarization for the tensor vacuum fluctuations, which could be seen by gravity wave detectors sensitive enough to probe
the primordial background, its degree of polarization and anisotropies. Looking at the 
anisotropy of the gravity waves linear polarization we identify the parity respecting and violating components of the effect. 
The imprint on the CMB temperature and polarization would be more elusive, since it averages to zero in the two-point functions, appearing only in  their cosmic variance or in fourth order correlators.
In contrast, spin zero correlations would have an effect on the two point function of the CMB temperature and polarization, enhancing the $BB$ component if they were anti-correlations. Such correlations represent an amplitude for the production of standing waves, as first envisaged by Grishchuk, and could also leave an interesting signature for gravity wave detectors. 
\end{abstract}

\keywords{cosmology}
\pacs{}

\maketitle


\section{Introduction}
The possibility that gravity might violate parity has been extensively studied in the past. This could happen for a number of
reasons, including Chern-Simons gravity~\cite{chern,chern2}, the renormalization of the gravitational coupling~\cite{leecj}, or the effects of quantum gravity (e.g.~\cite{Choi:1999zy, Freidel:2005sn, Takahashi:2009wc, joao1,joao2}). Parity violations could manifest themselves as 
asymmetries in the properties of right and left gravitons, or simply as a property of the vacuum and its fluctuations~\cite{joao1,joao2}. 
Parity-violating gravity is naturally inspired by grand unification and the fact that the standard model is chiral and maximally violates
parity~\cite{chern,chern2, Smolin:2007rx, Lue:1998mq}.  

In this paper we formally examine the general structure of gravitational violations of parity and identify a new phenomenon: the possible coupling between right and left graviton states. This could arise from the fact that they are not truly independent degrees of freedom
either because of a direct coupling between the two free Lagrangians, or more fundamentally as a result of the structure of the Hilbert space, and its inner product. At its most extreme it could manifest itself as oscillations between right and left gravitons. Less dramatically, it could correlate the vacuum fluctuations of right and left gravitons, and thus be imprinted in a primordial background, should it exist. 
It is the implications of the latter that we wish to explore in this paper.

We consider two distinct situations: correlations involving gravitons moving along the same direction, and along opposite directions. 
In the first case the effects could be both dramatic and subtle.  Foremost, there would be a degree of {\it linear} polarization in the gravitational wave background  (not to be confused with the linear polarization of the CMB photons, arising from Thomson scattering). 
The prospects for detecting a {\it circularly} polarized gravitational wave background have been evaluated, 
both using CMB polarization~\cite{leecj, Saito:2007kt, Gerbino:2016mqb} and direct detection~\cite{Seto:2007,Seto:2008, Kato:2015bye, Crowder:2012ik, Smith:2016jqs}.
As we will see, linear polarization in the gravitational waves would not be readily transcribed into a signature on the CMB photons, but would leave a rather subtle mark. In contrast, it would have a glaring imprint, should the the primordial gravity waves be directly detected  by an instrument sensitive to polarization and anisotropy. 

Specifically, linear polarization of primordial gravitational waves would introduce a degree of anisotropy in each realization. One could then evaluate their Stokes' parameters using a formalism similar to that used for light, but with spin 4 quantities, instead of spin 2. An E and B modes could be identified, in perfect analogy with electromagnetic waves. By evaluating the (isotropic) quadratic correlators we would select parity abiding  (EE and BB) and violating (EB) components. Their detection would present a formidable task for the new field of gravitational wave 
detection. 

The situation is somewhat different regarding correlations between right and left gravitons moving in opposite directions. 
Such correlations have spin 0 and represent an amplitude for the production of standing tensor wave, as initially proposed 
by Grishchuk~\cite{Grish}. They would
promptly show up in  the CMB  signature of tensor fluctuations, with a positive correlation suppressing and even erasing the BB component of the polarization, and an anti-correlation enhancing it. A spin 4 effect, in contrast, 
drops out of all CMB 2-point correlators. Thus, direct detection and CMB polarization can act as complementary probes.

In this paper we first consider the theoretical background of right-left graviton correlations (Section~\ref{theory}). We then 
investigate the most direct implications and lay down the general formalism (Sections~\ref{perfect} and~\ref{stokes}).
We then  evaluate the signature of the 2 types of corrections for a direct detection (Section~\ref{direct}) and via CMB power spectra (Section~\ref{CMB}). In a concluding Section we argue on the unexepcted nature of linear polarization in a gravitational wave background, and discuss the broad theoretical implications.

\section{Coupling right and left gravitons}\label{theory}
We start by considering possible theoretical motivations for
correlations between right and left tensor modes. Whilst we do not want
to wed ourselves to any particular model, these speculations provide interesting
background for the phenomenological models.

\subsection{Non-trivial  inner product in Hilbert space}

The chirality of vacuum fluctuations is usually expressed in the form of two power spectra:
\begin{eqnarray}\label{eq:prpl}
{\langle h_R(\mk)h^\star_R(\mk ')\rangle}&=&\delta(\mk-\mk')P_R(k),\nonumber\\
{\langle h_L(\mk)h^\star_L(\mk ')\rangle}&=&\delta(\mk-\mk')P_L(k),
\end{eqnarray}
with $P^-(k)=P_R(k) -  P_L(k)\neq 0$ resulting in parity violation. 
The brackets denote any kind of statistical average, but in the most common setting they are vacuum expectation values and represent the perturbative 2-point function of the theory. Whether or not the theory admits a trivial canonical quantization, these power spectra can be seen as norms of one-particle states: 
\be
{\langle 0| h_I(\mk)h^\star_I(\mk ')|0 \rangle}={\langle 1, \mk \,I |1, \mk'  \, I\rangle},
\ee
where $I=R,L$ (denoting right and left). 
The non-triviality of the normalization may be due to a deformation of the Hilbert space inner product, which in turn may result from a deformation of the measure in momentum space~\cite{Gubitosi:2015}, although this is not necessary. 
Parity violation may therefore be a manifestation of the asymmetric norms of right and left handed states. Moreover, 
whatever the origin of parity violation, it can be encoded in the inner product of the Hilbert space of the theory. 
The asymmetric norms of right and left states are then a way of expressing $P_R(k)\neq P_L(k)$.

Once we allow the matrix 
\be
M_{IJ}(k)={\langle 1, \mk \,I |1, \mk  \, J\rangle}
\ee
to fail to be proportional to the identity the question naturally arises: why should it be diagonal at all?
Specifically we could define a right-left correlator for vacuum fluctuations associated with gravitons moving along the same direction:
\be\label{PRL}
{\langle h_R(\mk)h^\star_L(\mk ')\rangle}=\delta(\mk-\mk')P_{RL}(\mk).
\ee
The reality conditions for $h_I ( {\mathbf x} )$ require $P_R$ and $P_L$ to be real, but $P_{RL}$ could in general be complex (the interpretation of its phase is left for Section~\ref{perfect}). The delta function in (\ref{PRL}) results from translational invariance, but
we refrain from dropping the direction of $\mk$ from the argument of $P_{RL}$ (an operation granted by isotropy).
The correlator $P_{RL}$ has spin 4, signalling an at least apparent violation of isotropy. In Section~\ref{stokes} we shall explore 
the full implications of this fact. 

A non-vanishing $P_{RL}$
could result from the fact that although the {\it deformed} inner product singles out the helicity states as independent degrees of freedom, when seen from the point of view of a theory with undeformed inner product these would appear non-orthogonal. 
It could also simply mean that the underlying orthogonal degrees of freedom are not the chiral states.
The correlations $P_{RL}$ may break parity, as we shall see, and this is somewhat surprising and related to the fact that $P_{RL}$ cannot in general be real. A parity transformation results in $P_{RL}\rightarrow P_{RL}^\star$, so that the real part of $P_{RL}$ is even, its imaginary part is odd. But it is important to stress that these possible violation, a priori, are independent of the more conventional parity violation encoded in $P^- (k)$. We could envisage a non-diagonal $M_{IJ}$ matrix with equal diagonal entries. 

\subsection{A more constrained model}

Taking cue from the phenomenon of neutrino oscillations (see Sec. 14 of \cite{Olive:2016xmw} for a review) we could consider a situation where ``conventional'' parity violation ($P^-\neq 0$)  and R-L correlations ($P_{RL}\neq 0$) are related because the helicity states are rotated from an underlying basis (denoted by indices $\tilde I$) which is indeed orthogonal with respect to the undeformed inner product. If $\theta$ is the ``mixing angle", 
and $M_{\tilde I\tilde J}=A^2\,  {\rm diag} (1+\epsilon,1-\epsilon)$, we have:
\bea
M_{RR}&=& A^2(1+ \ep\cos 2\theta)\\
M_{LL}&=& A^2( 1- \ep\cos 2\theta)\\
M_{RL}&=& A^2\frac{\ep}{2}\sin 2\theta.
\eea
In such a setting, the graviton hand-shake would only be non-vanishing when there is conventional parity violation ($\epsilon\neq 0$, and so
$P^-\neq 0$ except when $\theta=\pi/4$). However,
it would be possible to have parity violations without RL correlations, when $\theta=0$. The two phenomena can be parameterised by:
\bea
\frac{P_R-P_L}{P_R+P_L}&=&=\epsilon\cos 2\theta\\
\frac{\langle h_R|h_L\rangle}{P_R+P_L}&=&\frac{\ep}{2}\sin 2\theta\, .
\eea

As mentioned in the Introduction this sort of rotation could also arise at the level of the classical free Lagrangian, through 
the mass matrix, should  the graviton be massive (see~\cite{oscillations}). It could also involve massless gravitons via the kinetic terms in a field representation pegged down by the simplicity of the interactions. Such a rotation would be very different from 
the one considered in this paper, and it would result in oscillations between right and left gravitons, very similar to the flavour oscillations of neutrinos. 

\subsection{Coupling of chiral modes moving with opposite directions} \label{OppositeK}

The correlator $P_{RL}$ has spin 4, and so violates angular momentum conservation, at least apparently. 
This could be blamed on the fact that the actual conservation law is deformed, so that
its expression in terms of conventional quantities and addition rules results in an apparent violation. If this is the case, the same could happen regarding linear momentum and translational invariance \cite{AmelinoCamelia:1999pm, Gubitosi:2015}. The two types of
violations could occur concurrently or independently. 

One particularly interesting case, in a sense ``orthogonal'' to the one discussed so far,  
involves coupling between R and L gravitons moving in opposite directions:
\be\label{ARL}
{\langle h_R(\mk)h^\star_L(\mk ')\rangle}=\delta(\mk+\mk')A_{RL}(k).
\ee
This correlator has spin zero, and so does not entail any violation of angular momentum conservation, apparent or otherwise. Instead
it signals an apparent maximal violation of linear momentum conservation. 

As above, the deformation behind this correlator could result from the structure of the inner product of the Hilbert space, in 
particular regarding its component:
\be
{\langle 1, -\mk R | 1, \mk  L \rangle}\neq 0.
\ee
But more specific to this case it could arise from a deformation of momentum space related to the definition of
antipode of a vector in a curved momentum space \cite{AmelinoCamelia:1999pm, Arzano:2016egk}. In relation to this, it could result from an ambiguity in deformed field theories \cite{Arzano:2007ef}. 
The $\mk$ mode contains both the R/L graviton positive frequency moving in 
direction $\mk$ and the L/R graviton negative frequency moving in the $\-\mk$ direction
(note that the polarization tensors satisfy $\epsilon_{ij}^R(\mk)=\epsilon_{ij}^L(-\mk)$ implying the swap stated).
Curved momentum space could result in a  ``Bogolubov'' transformation between these modes, coupling 
R and L modes moving in opposite direction. 

In (\ref{ARL}) we have assumed isotropy, thereby rendering $A_{RL}$ a function of $k$ and not $\vk$. An interchange between R and L in the correlator can also be accomplished by a rotation by 180 degrees. Such an interchange would cause $A_{RL}\rightarrow A_{RL}^\star$, so isotropy forces $A_{RL}$  to be real. The correlator (\ref{ARL}) is intrinsically parity abiding.

As a test case we might consider the effect of 
perfect correlation (or anticorrelation) between modes moving in opposite directions. This would result in standing waves.
Indeed if $h_R(\vk)=h_L(-\vk)$ 
(corresponding to perfect correlation) the two modes for the headless direction $\pm\vk$ could be combined to produce:
\be\label{stand}
h(\vx)=h_R(\vk)[\cos(k\eta)\cos(\vk\cdot \vx)]
\ee
i.e. a circularly polarized standing wave. The polarization would count as R if we point the propagation to the right, L the other way. 
However, given that the wave is not propagating as such, R and L are not physically meaningful labels. The mode is chiral, though, since the polarization could be rotating in the opposite direction (i.e. we could have a $h_L(\vk)$ amplitude in (\ref{stand})). The two chiral modes can be converted into each other by a 180 degree rotation, as we have seen. An anti-correlation would result in a similar standing wave shifted by a quarter of a period. 
Regardless of what one thinks about the origin of these correlations, this is what was envisaged by Grishchuk as the outcome of a ``correct
quantum calculation of inflationary fluctuations"~\cite{Grish}. 

Any correlation of this type with a value between zero and the extreme case of perfect (anti)correlation would result in a superposition of travelling waves and standing waves.

\section{A test case: perfect coupling of chiral modes}\label{perfect}
In order to gain intuition on correlations between right  and left  gravitational waves moving along the same $\mk$, we first consider the extreme case of perfect correlation, i.e. every time a right wave is produced in a given direction and with a given frequency, a twin left one is generated, with the same amplitude {\it and phase}. From 
\be
h_{R/L}=\frac{h_+ \mp i h_\times}{\sqrt{2}}
\ee
and 
\bea
h_+&=&\frac{h_R+h_L}{\sqrt{2}}\nonumber \\
h_\times&=&\frac{h_L-h_R}{\sqrt{2} i}
\eea
(where linear polarizations are denoted by $+$ and $\times$)
we see that $h_R=h_L$  implies that only $+$  linearly polarized modes are produced. 
Such a fact breaks isotropy, since $+$ and $\times$ transform under a rotation around $\mk$
according to 
\bea
h_+&\rightarrow& \cos (2\phi_0) h_+ +\sin (2\phi_0) h_\times\\
h_\times&\rightarrow& -\sin (2\phi_0) h_+ +\cos (2\phi_0) h_\times
\eea
if $\phi\rightarrow \phi +\phi_0$. Thus, pure $+$ polarization can only be true in one frame, breaking isotropy. 
This can be traced to the fact that the amplitudes of R and L modes are rotationally invariant, but their phases are not:
\be
h_{R/L}\rightarrow e^{\pm 2i\phi_0}h_{R/L}.
\ee
Since the phases of R and L modes transform differently, the requirement that R and L have the same phase can only be true in one frame. 
This is the frame where pure $+$ polarization is observed, and in which $P_{RL}$ is real. In a general frame $P_{RL}$ is complex and its phase carries information on the preferred direction selected by the correlations. 

This extreme example illustrates the fact that the correlator $P_{RL}$ is not invariant under rotations, but transforms as:
\be
P_{RL}\rightarrow P_{RL}e^{4i\phi_0},
\ee
that is, it has spin 4.
We saw above that $P_{RL}$ does not need to be real, but we can now add to this. 
Even if
we posited that for a given theory $P_{RL}$ is real, 
this could only be true in one frame, and undone with a rotation.
Right-left correlations invariably break isotropy in each realization, and the phase of $P_{RL}$ carries the information on the preferred direction selected. 

Of course, we can postulate that the preferred direction selected by each realization is random within an ensemble,
so that there must be an isotropic description of the phenomenon. This amounts to stating that  in a concrete frame 
$P_{RL}=\sqrt{P_R P_L} e^{i\theta}$ with $\theta$ then given a uniform distribution. However, this averages 
$P_{RL}$ to zero, so that perfect isotropic correlation between R and L can only be encoded in 
${\langle h_R h_L^\star h_L^\star h_R\rangle}=P_R P_L$, i.e. involve fourth order correlators (and non-Gaussianity). As we
shall see, this will be very relevant  in the investigation of the effects on the CMB.

\section{The general formalism}\label{stokes}
The example of the previous Section shows that perfect correlation between R and L marks the onset of pure linear polarization, 
in such a way that in the frame where $P_{RL}$ is real we observe pure $+$ polarization. In all other frames 
$P_{RL}$ has a phase, carrying information of the direction of the pure $+$ polarization. This is the physical meaning
of the phase of $P_{RL}$.  

It turns out that even in less extreme cases, correlations between R and L gravitons translate into a degree of linear polarization.
This can be quantified by the Stokes parameters, with the quantity $P_{RL}$ already playing an important role in that setting. 
The formalism for gravitational waves polarization is similar to that used for EM waves~\cite{Thorne, Seto:2007,Seto:2008}.
Using this formalism we can map
the 2 by 2 self-adjoint matrix of graviton correlators formed by $P_R$, $P_L$ and $P_{RL}$ into the ``Stokes parameters'' 
$I$, $V$, $Q$ and $U$ according to:
\bea
I&=&P_R+P_L\\
V&=&P_R-P_L\\
Q&=&2\Re P_{RL}\\
U&=&2\Im P_{RL},
\eea
or:
\bea 
P_{R}&=&\frac{I + V}{2}\\
P_{L}&=&\frac{I - V}{2}\\
P_{RL}&=&\frac{Q+i U}{2}.
\eea
Parameter $I$ is the overall power spectrum and $V$ is the degree of circular polarization. Both have spin 0. 
Parameters $Q$ and $U$ measure the linear polarization, and in fact $P_{RL}$ already constitutes their complex combination introduced in the standard formalism because it is easy to study, since it has definite spin.  
As mentioned above, $P_{RL}$ has spin 4 (for a spin 2 graviton), just as for photons 
(with spin 1) the linear polarization $(Q+iU)/2$ has spin 2. Thus
$P_{RL}$ can never be isotropic (its multipole expansion must start at $\ell=4$), which is why we kept the vector nature of $\mk$ in its argument in Eq.~(\ref{PRL}).

As mentioned in the previous Section the fact that each realization is anisotropic does not preclude the creation of an isotropic ensemble, but its properties must be encoded in the 4-point function of the fields ($h$ for gravity, the $E$ and $B$ fields for light). This is the subject of next Section. For each realisation the presence of any degree of linear polarization inevitably picks up a preferred direction. 
The only way in which a wave can avoid doing this is either to have no polarization at all, or be purely circularly polarized. Any admixture of linear polarization (i.e. any elliptical polarization) selects a preferred direction in the plane perpendicular to the direction of propagation. 

This can be seen in the $+\times$ basis, too, with correlators:
\bea
P_+={\langle |h_+|^2\rangle}&=&\frac{P_R+P_L}{2}+\Re P_{RL}=\frac{I+Q}{2}\nonumber\\
P_\times={\langle |h_\times|^2\rangle}&=&\frac{P_R+P_L}{2}-\Re P_{RL}=\frac{I-Q}{2}\nonumber\\
P_{+\times}={\langle h_+h^\star_\times \rangle}&=&\frac{P_R-P_L}{2i}-\Im P_{RL}= -\frac{U+iV}{2}.\nonumber
\eea
It is not enough to require  $P_+ = P_\times$ for the polarization of a given wave to be isotropic,
because one must still consider $P_{+\times}$. However, an imaginary $P_{+\times}$ does not break isotropy 
because it is associated with purely circular polarization. Indeed:
\bea
I&=&P_++P_\times\\
Q&=&P_+ -P_\times \\
 U&=&- 2 \Re P_{+\times}\\
V&=& - 2\Im P_{+\times},
\eea
clarifying this matter. The choice of basis is of course a matter of taste, but we shall use the R-L basis for the rest of this paper.

\section{Detection by gravity wave experiments}\label{direct}
It turns out that the best way to see R-L correlations in gravitational waves {\it moving in the same direction} is by direct detection,
with gravity wave experiments sensitive to polarization~\cite{gravback1,gravback2}. Indeed, the observation of any degree of linear
polarization in the gravitational wave background (as opposed to a degree of circular polarization) would be a direct detection of $P_{RL}$. In addition, should the experiment mapping abilities allow it,  we would be able to construct the isotropic quadratic correlators of $P_{RL}$ (assuming the underlying ensemble is isotropic, even if each realization is not). These would be quartic in the fields $h$. The formalism closely mimics that used for the CMB light~\cite{selzal}, bearing in mind that the linear polarization $Q+iU$ is now spin 4 instead of 2.

As in~\cite{Seto:2007,Seto:2008}, let us expand $P_{RL}=(Q+iU)/2$ seen at a given point into spin 4 spherical harmonics:
\be
P_{RL}(f,\vn)=\int df \sum_{\ell=4}^{\infty}\sum_{m=-\ell}^\ell \, _4b_{\ell m}(f)  \, _4Y_{\ell m}(\vn)
\ee
where $\vk= f \vn$, with $\vn$ is the direction of observation and $f$ the frequency. Given the spin 4 nature of $P_{RL}$
the angular expansion starts at $\ell=4$. The $P_{RL}^\star$ can be similarly expanded in spin $-4$ harmonics, with coefficients ${}_{-4}b_{\ell m}={}_4b^\star_{\ell -m}$.
As in~\cite{selzal} we can then extract from ${}_{\pm 4}b_{\ell m}$  the parity even and odd components, noting that under a parity transformation $\vn\rightarrow -\vn$, R and L interchange, $P_{RL}\rightarrow P_{RL}^\star$, 
and consequently ${}_{\pm 4}b_{\ell m}$ interchange too. Hence the parity even and odd components (the ``electric'' and ``magnetic '' components) have coefficients:
\bea
b^E_{\ell m}&=&-\frac{{}_4b_{\ell m} + {}_{-4}b_{\ell m}}{2} \\
b^B_{\ell m}&=&i\frac{{}_4b_{\ell m} -  {}_{-4}b_{\ell m}}{2} 
\eea
(where the factors of $-$ and $i$ are conventional; several different conventions are in use). 
As for the CMB~\cite{selzal} we can set up correlators between the various $b^I_{\ell m}$ (with $I=E,B$),
with power spectra defined as:
\be
{\cal G}_\ell^{IJ}=\frac{1}{2\ell +1}\sum_m {\langle b^I_{\ell m} b^J_{\ell m}\rangle}
\ee
where we used ${\cal G}_\ell$ here to avoid confusion with the CMB ${\cal C}_\ell$ introduced in the next Section.  We have 3 such power-spectra: $EE$, $BB$ and $EB$. The first two are parity invariant, whereas the third signals parity violation at the level of correlations between R and L (not to be confused with the more conventional $P_R\neq P_L$, leading to parameter $V$ even before we consider another level of correlators). We see that by looking at the anisotropy in the gravity wave backgound we can unearth a somewhat counterintuitive 
phenomenon. Right and left gravity correlations can in fact break parity invariance. The effect is fourth order in the 
amplitudes and does not have a monopole, but it could in principle be present.

One may wonder in what sense can we take averages twice for gravitational waves, once leading to $P_{RL}$ another to 
${\cal G}_{\ell}^{IJ}$. In fact the analogy with CMB polarization is only formal, and the two are quite different in this respect.
In the case of the polarization of the CMB the averaging leading from EM fields to Stokes parameters refers to an average
over several coherence lengths of non-coherent light; the averaging behind the ${\cal C}_\ell^{IJ}$
is then taken over a cosmological ensemble. The gravitational wave background is perfectly coherent and the two tiers of 
averages both concern a cosmological ensemble.  Correlations between R and L break isotropy on a realization by realization basis, but not in terms of the whole ensemble. If we want to isotropize $P_{RL}$ it must be zero, because the average of a random phase $e^{i\theta}$ is zero. Instead we may conditionalize the distribution to the direction of a given linear polarization, and interpret the first average as over these sub-ensembles. By marginalizing over the direction we isotropize the ensemble, with the second averaging (leading to the
${\cal G}_\ell^{IJ}$) corresponding to this process.

As with the CMB we can produce spin zero versions of the $Q$ and $U$ by 
acting 4 times on the spin-4 fields with spin raising and lowering
operators $\dup$ and $\ddn$ \cite{selzal}
\begin{eqnarray}
  \tilde E_G({\bf \hat n}) &=& - \frac{1}{2}\left[ \ddn^4P_{RL}({\bf \hat n})
  + \dup^4P_{RL}^\star({\bf \hat n})\right],\nonumber\\
 \tilde B_G({\bf \hat n}) &=& \frac{i}{2}\left[ \ddn^4P_{RL}({\bf \hat n})
  - \dup^4P_{RL}^\star ({\bf \hat n})\right].
\end{eqnarray}
However, the power spectra are not built from these quantities.

\section{Effects upon the CMB}\label{CMB}
CMB power spectra (both temperature and polarization) include contributions from tensor modes, i.e. primordial gravity waves. It was  shown in \cite{leecj}, that parity violating fluctuations (with $P_{R}\neq P_{L}$) would show up in the TB and EB cross-correlators. So one could wonder whether RL correlations could also be seen in the CMB, both in the case of waves moving along and opposite directions.

\subsection{Negative result for gravitons moving along the same direction}
It turns out that for gravitons moving along the same direction this is not possible at the level of the power spectra. The $P_{RL}$ correlator has spin four and so cannot appear in the $C_{\ell}$'s, which are spin zero. By contrast, the effects of $P_R\neq P_L$  can be seen in the CMB because the $P_{R}$ and $P_{L}$ correlators are spin zero, as are the  $I$ and $V$ Stokes parameters. The following explicit calculation shows how the $P_{RL}$ contribution drops out of the power spectra.

Following \cite{selzal}, the contribution to the power spectra from tensor modes is given by:
\be
\mathcal C_{\ell}^{XY}=\frac{1}{2\ell+1} \sum_{m}\langle a_{\ell m,X}^{(T)*}a_{\ell m,Y}^{(T)}\rangle
\ee
where $X,Y=T,E,B$, the superscript $(T)$ denotes tensor modes and
\bea
a_{\ell m,T}^{(T)}&=& \int d\Omega \, Y^{*}_{\ell m}(\hat{\vn})\Delta_{T}^{(T)}(\hat{\vn})\nonumber\\
a_{\ell m,E,B}^{(T)}&=&  \sqrt{\frac{(\ell-2)!}{(\ell+2)!}} \int d\Omega \, Y^{*}_{\ell m}(\hat{\vn})\Delta_{\tilde E, \tilde B}^{(T)}(\hat{\vn})\,.
\eea
$\Delta_{X}^{(T)}(\hat \vn)\equiv \int d^{3}\vk \Delta_{X}^{(T)}(\hat \vn,\tau_{0},\vk)$ are the anisotropies generated by tensor modes. They are given by combinations of R and L gravity waves, multiplied by quantities that depend on the integral over the line of sight of the source functions, which we schematically call $\mathcal{S}_{T,P}$  \cite{selzal}: 
\bea
\Delta_{T}^{(T)}(\hat \vn,\tau_{0},\vk)&=&\left[e^{2 i \phi}h_{R}(\vk)+e^{-2 i \phi}h_{L}(\vk)\right]\mathcal{S}_{T}\nonumber\\
\Delta_{\tilde E}^{(T)}(\hat \vn,\tau_{0},\vk)&=&\left[e^{2 i \phi}h_{R}(\vk)+e^{-2 i \phi}h_{L}(\vk)\right] \mathcal{S}_{P}\nonumber\\
\Delta_{\tilde B}^{(T)}(\hat \vn,\tau_{0},\vk)&=&\left[e^{2 i \phi}h_{R}(\vk)-e^{-2 i \phi}h_{L}(\vk)\right]\mathcal{S}_{P}\label{eq:TensorPerturbations}
\eea
The angle $\phi$ is the azimuthal angle of $\hat \vn$ in the reference frame where $\vk ||\hat \vz$ \cite{kosowsky}.

As an example, 
say we want to compute the contribution of tensor modes to the $C_{\ell}^{TE}$ spectrum:
\begin{widetext}
\bea
\mathcal C_{\ell}^{TE}&=&\mathcal N(\ell) \;\sum_{m} \langle \left( \int d\Omega\, Y^{*}_{\ell m}(\hat \vn)  \int d^{3}\vk\,  \left[e^{2 i \phi}h_{R}(\vk)+e^{-2 i \phi}h_{L}(\vk)\right]\mathcal S_{T}  \right)^{*}\cdot \nonumber\\
&&\cdot\left( \int d\Omega' \,Y^{*}_{\ell m}(\hat \vn')  \int d^{3}\vk' \,\left[e^{2 i \phi'}h_{R}(\vk)+e^{-2 i \phi'}h_{L}(\vk)\right] \mathcal{S}_{P}   \right)\rangle\nonumber\\\label{eq:CTE}
\eea

\end{widetext}
with $\mathcal N(\ell)= \frac{1}{2\ell+1} \left(\frac{(\ell-2)!}{(\ell+2)!}\right)^{1/2} $. The contribution from the RL correlator comes from the term proportional to
\bea&& e^{2 i (\phi'+\phi)}\langle h_{L}^{*}(\vec k)h_{R}(\vec k')\rangle +e^{- 2 i (\phi'+\phi)} \langle h_{R}^{*}(\vec k) h_{L}(\vec k')\rangle \nonumber\\
&&=\left[ e^{2 i (\phi'+\phi)}P_{RL}(k)+e^{- 2 i (\phi'+\phi)}P_{RL}^{*}(k)\right]\delta^{(3)}(\vk-\vk ') \nonumber\eea
Expanding the spherical harmonics in terms of Legendre polynomials, $Y_{\ell m}(\hat\vn)=\left(\frac{(2\ell+1)(\ell-m)!}{4\pi (\ell+m)!}\right)^{1/2} P_{\ell}^{m}(\mu)e^{im\phi}$, one can see that the $P_{RL}$ and $P_{RL}^{*}$ contributions are respectively proportional to the angular integrals \footnote{The integrals over the angular volume are decomposed as: $\int d\Omega=\int_{0}^{2\pi} d\phi\int_{-1}^{1} d\cos\theta$}:
\be
\int d\phi \int d\phi' e^{i m \phi} e^{-im\phi'} e^{2 i (\phi'+\phi)}
\ee  
and
\be
\int d\phi \int d\phi' e^{i m \phi} e^{-im\phi'} e^{-2 i (\phi'+\phi)}
\ee  
These are clearly both vanishing, given that $\int d\phi \,e^{i (a-b) \phi}\propto \delta_{ab}$. Note that this vanishing result is exactly due to the  different spin properties of the power spectrum and the RL correlator.
The same thing happens, for the same reasons, for all the other ${\cal C}^{XY}$.
In their standard calculation one often assumes that $P_{RL}$ is zero. In fact this assumption is not necessary. The standard result 
does not change even if $P_{RL}\neq 0$.

\subsection{Non-Gaussianity and higher-order correlators}
Naturally the effect may show up in higher order correlators, with a degree of non-Gaussianity implied, so that Wick's theorem does not apply. It is often the case that a Gaussian anisotropic process can be isotropized by assigning a uniform distribution $P(\vn)=1$ for its preferred axis 
$\vn$ (either because we do not know it, or because this is intrinsically the case). The conditional distribution for a given observable $P({\cal O}|\vn)$ is then Gaussian, but the marginalized distribution:
\be
P({\cal O})=\int d\vn\, P({\cal O}|\vn)P(\vn)
\ee
is not. The full isotropic ensemble is therefore broken into a set of isotropic sub-ensembles~\cite{FerMag}. In our case, averages taken over the anisotropic sub-ensemble lead to second order spin 4 correlators. A second level of averages, over the preferred direction, lead to fourth order isotropic correlators. This is just the two-tier process of averages referred to in Section~\ref{direct}.

Right-left correlations for gravitons moving along the same direction would then show up in the fourth order correlators of CMB quantities. 
By implication it would also affect the cosmic variances for CMB power spectra estimators, which would be anomalous with respect to their Gaussian values. In effect, even if the effect on the average or theoretical ${\cal C}_\ell^{XY}$ is zero, anomalies would appear in most realizations. 

We leave to a future publication a full investigation of these issues, but it would appear that the most significant marks of these correlations would {\it not} be left in the CMB.

\subsection{Correlations for gravitons moving along opposite directions}
In contrast, a correlator between  chiral modes moving in opposite directions (as proposed in \ref{OppositeK}) 
would show up readily in the CMB power spectra, since such correlation is a spin-zero quantity.
This can be proved by returning to Eq. (\ref{eq:CTE}) and noting that now the contribution from the RL correlator comes from
\bea&& e^{2 i (\phi'+\phi)}\langle h_{L}^{*}(\vec k)h_{R}(\vec k')\rangle +e^{- 2 i (\phi'+\phi)} \langle h_{R}^{*}(\vec k) h_{L}(\vec k')\rangle \nonumber\\
&&=\left[ e^{2 i (\phi-\phi')}A_{RL}(k)+e^{- 2 i (\phi-\phi')}A_{RL}^{*}(k)\right]\delta^{(3)}(\vk+\vk ') .\nonumber\eea
Since the angle $\phi$ ($\phi'$) is the azimuthal angle of $\hat \vn$ ($\hat \vn'$) in the reference frame where $\vk||\hat \vz$ ($\vk' ||\hat \vz$), flipping the direction of $\vk$ ($\vk'$) results in a change of sign of $\phi$ ($\phi'$).
Thus, the contributions from $A_{RL}$ and $A_{RL}^{*}$ are respectively proportional to:
\be
\int d\phi \int d\phi' e^{i m \phi} e^{-im\phi'} e^{2 i (\phi-\phi')}\propto \delta _{m,-2}.
\ee  
and
\be
\int d\phi \int d\phi' e^{i m \phi} e^{-im\phi'} e^{-2 i (\phi-\phi')}\propto \delta _{m,2}.
\ee  
Once this angular contribution is correctly taken into account, one can follow the same steps as in \cite{selzal} to find the tensor contribution to the power spectra : 
\bea
\mathcal C_{\ell}^{XY}&=& (4\pi)^{2}\int k^{2} dk \left[  P_{R}+P_{L}+ 2 A_{RL} \right] \Delta_{X}^{(T)}\Delta_{Y}^{(T)} \nonumber\\
\mathcal C_{\ell}^{XB}&=& (4\pi)^{2}\int k^{2} dk \left[  P_{R}-P_{L} \right] \Delta_{X}^{(T)}\Delta_{B}^{(T)}\nonumber\\
\mathcal C_{\ell}^{BB}&=& (4\pi)^{2}\int k^{2} dk \left[  P_{R}+P_{L}- 2  A_{RL} \right] \left[\Delta_{B}^{(T)}\right]^{2}\nonumber\\\label{eq:ARLspectra}
\eea
where $X,Y=T,E$ and $\Delta_{X,B}^{(T)}\equiv\Delta_{X,B}^{(T)}(k)$ are defined in \cite{selzal}, eq. (30). 
We have used the fact that $A_{RL}$ must be real, as consequence of isotropy (assumed in the calculation). Even if the assumption of isotropy was dropped, making $A_{RL}$ complex, only the real part of $A_{RL}$ would contribute to the power spectra.
The standard power spectra sourced by the primordial spectrum of gravity waves $P_{h}$ are obtained setting $P_{R}=P_{L}=P_{h}/2$ and $A_{RL}=0$.

We see that $A_{RL}$ has opposite effects on the BB spectrum with respect to TT, TE and EE spectra, leading to suppression of the spectrum in one case and enhancement of the spectrum in the other, if $ A_{RL}>0$. 
In particular, an anti-correlation would enhance the amplitude of the BB spectrum with respect to the other parity abiding spectra. 
Since parity cannot be violated by this effect, the parity odd components TB and EB do not receive any contribution. 

In the extreme case of perfect correlation, corresponding to one of the standing waves modes, the BB signature of the tensors would be perfectly erased. In contrast an anti-correlation, corresponding to the other standing wave mode, would double the value of the BB component. In our setting we violate translational invariance, but as in the case of isotropy and spin 4 correlations, this could be 
restored by considering an ensemble with sub-ensembles corresponding to a given value of $A_{RL}$ subject to a uniformly distributed translational shift. The contributions to the 2-point functions would then average to zero, but the effect would show up realization by realization, or in the cosmic variance of the estimators. In work in preparation we will investigate this matter, and compare with the results in~\cite{Grish-poln}.

\section{Discussion}
In this paper we investigated the effects of correlations between right and left handed gravitons moving along the same direction, and moving in opposite directions. We motivated these correlations with deformations of the inner product in Hilbert space, as well as deformations to the usual conservation laws and/or addition rules, as suggested by some quantum gravity phenomenologies. Depending on whether the correlations affect gravitons moving along the same or opposite directions, one can relate the deformations to apparent violations to angular or linear momentum conservation, respectively. The observational effects are quite different in the two cases.

Correlations involving gravitons with the same frequency and direction would manifest themselves in a degree of linear polarization in the gravitational wave background. This would be seen in a direct detection. We invoked the theory of gravity wave polarization, showing how the correlator can be directly used to define Stokes parameters, and combined into isotropic measures, separating the parity even and odd components. The latter are fourth order in the (measured) amplitudes and probably difficult to measure, but would contain the most information. 

Linear polarization of gravity waves would be a unique signature of quantum gravity, since it is difficult to account for it in any other way. Linear polarization of light is ubiquitous and arises from reflection, but this is not an option for gravity waves, which barely interact with anything else, let alone be reflected. 
Obviously, the astrophysical ``foregrounds'' are polarized, but they are time dependent and so eminently ``subtractable''. 
The observation of linear polarization in the gravity wave background would relate directly to a graviton hand-shake, for gravitons moving in the same direction.  The only assumption is that it derives from vacuum fluctuations (see~\cite{string-gas} for a counter-example to this assumption). 
By contrast, the effect of linearly polarized gravity waves on the CMB polarization would be very subtle. The effects drop out of the quadratic CMB power spectra, manifesting themselves only in the variance of its estimators, or in fourth order polarization correlators. A non-Gaussian effect on these is predicted. 

The situation is entirely different regarding right-left correlations for gravity waves moving in opposite directions. These would show up in the two point correlators for CMB temperature and polarization. A positive correlation would suppress the BB power with respect to TT, TE and EE. An anti-correlation would have the opposite effect \footnote{Since $A_{RL}$ enters in the BB power spectrum in the same way as the usual tensor contributions, $P_R+P_L$ (see eq. \eqref{eq:ARLspectra}), the issue of disentangling the constraints on $A_{RL}$ and the ones on the tensor-to-scalar ratio $r$ arises. The exact level of degeneracy depends however on the dependence of $A_{RL}$ on $k$ as compared to the one of $P_R + P_L$.}. Given that such a graviton correlation cannot violate parity (unlike correlations involving gravitons moving along the same direction), the effect does not contribute to TB and EB correlators. Such correlations correspond to an amplitude for the existence of standing waves, as first proposed in~\cite{Grish}. The imprint on gravity wave detection has been the subject of past investigations, but we hope to research it further.

We close with a clarification. 
In spite of the obvious formal parallel between the polarization formalisms for light and for gravity, there are significant differences. The polarization coherence matrix is defined for the CMB light because it is incoherent. 
By contrast the primordial gravity waves are like a laser, since the waves have perfect time phase coherence (as opposed to the spatial phases, which are uniformly distributed for a Gaussian random field). Therefore the averaging processes leading from fields to polarization, 
and from polarization Stokes parameters to power spectra is different in the two cases. 
For gravity waves both processes are a cosmological ensemble average, as described in Section~\ref{direct}. For CMB light only the second is a cosmological ensemble average, with the first being an average over coherence lengths of light.

Also, for gravity waves we measure directly the fields $h$,  and these are considered first order, so that the polarization is second order, and its power spectra are fourth order. For CMB light we measure directly the polarization parameters, which are considered first order. 
The EM field waves are never measured per se and would be ``order 1/2'', whereas the polarization power spectra are second order.
For these reasons the two types of polarization do not interact  in a one-to-one way. 
Overall, CMB polarization and gravity wave background polarization appear to be complementary probes of the Early Universe.

\section{Acknowledgments}
We thank Giovanni Amelino-Camelia, Michele Arzano and Carlo Contaldi for discussions related to this paper, and 
acknowledge support from the John Templeton Foundation. JM was also funded by an STFC consolidated grant 
and the Leverhulme Trust.

\end{document}